\def\apjl{Astrophys. J. Lett.}
\def\apjs{Astrophys.J.Suppl.}

\def\aap{Astron. Astrophys.}

\newcommand{\ra}{\;\raise1.0pt\hbox{$'$}\hskip-6pt\partial\;}
\newcommand{\lo}{\;\overline{\raise1.0pt\hbox{$'$}\hskip-6pt\partial}\;}

\documentclass[traditabstract]{aa} 
\usepackage{graphicx}
\usepackage{txfonts}
\usepackage{natbib}

\begin{document}
\title{CMB E/B decomposition of incomplete sky: a pixel space approach} \subtitle{}
\author{Jaiseung Kim\thanks{jkim@nbi.dk} \and Pavel Naselsky}
\institute{Niels Bohr Institute \& Discovery Center, Blegdamsvej 17, DK-2100 Copenhagen, Denmark}

\abstract{
CMB polarization signal may be decomposed into gradient-like (E) and curl-like (B) mode.
We have investigated E/B decomposition in pixel space. We find E/B mixing due to incomplete sky is localized in pixel-space, and negligible in the regions far away from the masked area. By estimating the expected local leakage power, we have diagnosed ambiguous pixels. Our criteria for ambiguous pixels (i.e. $r_c$) is associated with the tensor-to-scalar ratio of B mode power spectrum, which the leakage power is comparable to.
By setting $r_c$ to a lower value, we may reduce leakage level, but reduce sky fraction at the same time.
Therefore, we have solved $\partial \Delta C_l/\partial r_c=0$, and obtained the optimal $r_c$, which minimizes the estimation uncertainty, given a foreground mask and noise level. We have applied our method to a simulated map blocked by a foreground (diffuse + point source) mask. Our simulation shows leakage power is smaller than primordial (i.e. unlensed) B mode power spectrum of tensor-to-scalar ratio $r\sim 1\times10^{-3}$ at wide range of multipoles ($50\lesssim l \lesssim 2000$), while allowing us to retain sky fraction $\sim 0.48$.
}
\maketitle

\section{Introduction}
Over the past years, CMB polarization has been measured by several experiments and is being measured by the Planck surveyor \citep{DASI:data,DASI:instrument,DASI:I,DASI:II,DASI:III,DASI:3yr,QUaD1,QUaD2,QUaD:instrument,QUaD_improved, Planck_bluebook}.
CMB polarization pattern may be considered as the sum of gradient-like E mode and curl-like B mode \citep{Seljak-Zaldarriaga:Polarization,Kamionkowski:Flm}.
In the standard model, B mode polarization is not produced by scalar perturbation, but solely by tensor perturbation. 
Therefore, measurement of B mode polarization makes it possible to probe the universe on the energy scale at inflationary period \citep{Kamionkowski:Flm,Seljak-Zaldarriaga:Polarization,Modern_Cosmology,Inflation,Foundations_Cosmology}. 
In most inflationary models, tensor-to-scalar ratio $r$ is much smaller than one, and the WMAP 7 year data imposes an upper bound on $r<0.36$ at $95\%$ confidence level \citep{WMAP7:powerspectra,WMAP7:Cosmology}.

Besides instrument noise, there are complications, which limits detectability of tensor perturbation. Imperfection in removing foreground and gravitational lensing imposes observational limit on tensor-scalar-ratio: $r\sim10^{-4}$ and $r\sim 3\times10^{-5}$ respectively \citep{B_lensing,B_mode_limit_foreground}. 
Due to the nature of the observation or heavy foreground contamination, reliable of estimation on CMB polarization signal is not available over a whole sky.
Incomplete sky coverage leads to E/B mixing, and very significantly limit our capacity to measure tensor perturbation as well \citep{Bunn:EB-Separation}.
Therefore, there have been various efforts to understand and reduce E/B mixing \citep{Kim:optimization,Kim:measuring_a2lm,Bunn:EB-Separation,EB_incomplete_sky,EB_harmonic,Smith:pseudo_EB}.

It is best to implement E/B decomposition in map space, since diffuse foregrounds and point sources are well-localized in map space, and their spatial information are known relatively better than other properties. In this paper, we investigate E/B decomposition in pixel space.
Our investigation shows that E/B mixing is highly localized in pixel space.
Therefore, we may reduce E/B mixing effectively by excluding the ambiguous pixels. We have applied our method to simulated maps partially blocked by a foreground (diffuse + point source) mask.  After excluding ambiguous pixels, we find that leakage power in retained pixels (sky fraction $\sim 0.48$) is smaller than primordial (i.e. unlensed) B mode power spectrum of tensor-to-scalar ratio $r\sim 1\times10^{-3}$ at wide range of multipoles ($50\la l \la 2000$).

The outline of this paper is as follows. 
In Sec. \ref{Stokes}, we discuss all-sky analysis of CMB polarization.
In Sec. \ref{pixel_filter}, we derive E/B decomposition in pixel space. 
In Sec. \ref{cutsky}, we discuss the application to cut sky, and the method to diagnose ambiguous pixels.
In Sec. \ref{simulation} and \ref{scale}, we present our simulation result.
In Section \ref{Discussion}, we summarize our investigation. 
In Appendix \ref{noise}, we discuss error analysis of pseudo $C_l$ estimation, and show interpixel noise correlation may be neglected.

\section{STOKES PARAMETERS}
\label{Stokes}
The state of polarization is described by Stokes parameter \citep{Kraus:Radio_Astronomy,Tools_Radio_Astronomy}.
Since Thompson scattering does not generate circular polarization, Stokes parameter Q and U are sufficient to describe CMB polarization \citep{Modern_Cosmology}. 
Stokes parameter $Q$ and $U$ transform under rotation of an angle $\psi$ on the plane perpendicular to direction $\mathbf {\hat n}$ \citep{Seljak-Zaldarriaga:Polarization,Zaldarriaga:Polarization_Exp}: 
\begin{eqnarray}
\label{Q'U'} (Q\pm \imath U)'(\mathbf {\hat n})=e^{\mp 2i\psi}(Q\pm \imath U)(\mathbf {\hat n}).
\end{eqnarray}
Therefore, all-sky Stokes parameters may be decomposed into spin $\pm2$ spherical harmonics \citep{Seljak-Zaldarriaga:Polarization} as follows:
\begin{eqnarray}
Q(\mathbf {\hat n})\pm i U(\mathbf {\hat n})&=&\sum_{l,m} a_{\pm2,lm}\;{}_{\pm2}Y_{lm}(\mathbf {\hat n}),\label{Q_lm+iU_lm} 
\end{eqnarray}
where the decomposition coefficients $a_{\pm2,lm}$ are obtained by:
\begin{eqnarray}
a_{\pm2,lm}=\int \left[Q(\mathbf {\hat n})\pm i U(\mathbf {\hat n})\right]\,{}_{\pm2}Y^*_{lm}(\mathbf {\hat n})\,\mathrm d \mathbf {\hat n}.\label{a2lm}
\end{eqnarray}
Though the quantity shown in Eq. \ref{Q_lm+iU_lm} has direct association with physical observables (i.e. Stokes parameters), rotational variance leads to computational complication. Therefore, two real scalar quantities, termed `E' and `B' mode, are often built out of $Q(\mathbf {\hat n})\pm i U(\mathbf {\hat n})$ \citep{Kamionkowski:Flm,Seljak-Zaldarriaga:Polarization}:
\begin{eqnarray}
E(\mathbf {\hat n})&=&-\frac{1}{2}[\lo^2 (Q(\hat {\mathbf n}) + i U(\hat {\mathbf n}))+\ra^2 (Q(\hat {\mathbf n})- i U(\hat {\mathbf n}))],\nonumber\\
&=&\sum_{lm} \sqrt{\frac{(l+2)!}{(l-2)!}}\,a_{E,lm}\,Y_{lm}(\mathbf {\hat n}),\label{E_map}
\end{eqnarray}
\begin{eqnarray}
B(\mathbf {\hat n})&=&\frac{\imath}{2}[\lo^2 (Q(\hat {\mathbf n}) + i U(\hat {\mathbf n}))-\ra^2 (Q(\hat {\mathbf n})- i U(\hat {\mathbf n}))]\nonumber,\\
&=&\sum_{lm} \sqrt{\frac{(l+2)!}{(l-2)!}}\,a_{B,lm}\,Y_{lm}(\mathbf {\hat n})\label{B_map},
\end{eqnarray}
where $\lo$ and $\ra$ refer to lowering and raising operator respectively \citep{Seljak-Zaldarriaga:Polarization}.
The explicit expression of $\lo$ and $\ra$ are given as follows \citep{Seljak-Zaldarriaga:Polarization}:
\begin{eqnarray*}
\ra {}_s f(\theta,\phi)&=&-\sin^s \theta\left[\frac{\partial}{\partial \theta} +\imath \csc\theta \frac{\partial}{\partial \phi}\right] \sin^{-s}\theta\;{}_s f(\theta,\phi),\\
\lo {}_s f(\theta,\phi)&=&-\sin^{-s} \theta\left[\frac{\partial}{\partial \theta} -\imath \csc\theta \frac{\partial}{\partial \phi}\right] \sin^{s}\theta\;{}_s f(\theta,\phi),
\end{eqnarray*}
where ${}_s f(\theta,\phi)$ is an arbitrary spin $s$ function. 
The decomposition coefficients of E and B mode are related to $a_{\pm 2,lm}$\citep{Seljak-Zaldarriaga:Polarization} as follows:
\begin{eqnarray}
a_{E,lm}&=&-(a_{2,lm}+ a_{-2,lm})/2,\label{alm_E}\\
a_{B,lm}&=&i(a_{2,lm} - a_{-2,lm})/2.\label{alm_B}
\end{eqnarray}

\begin{figure}
\includegraphics[scale=.5]{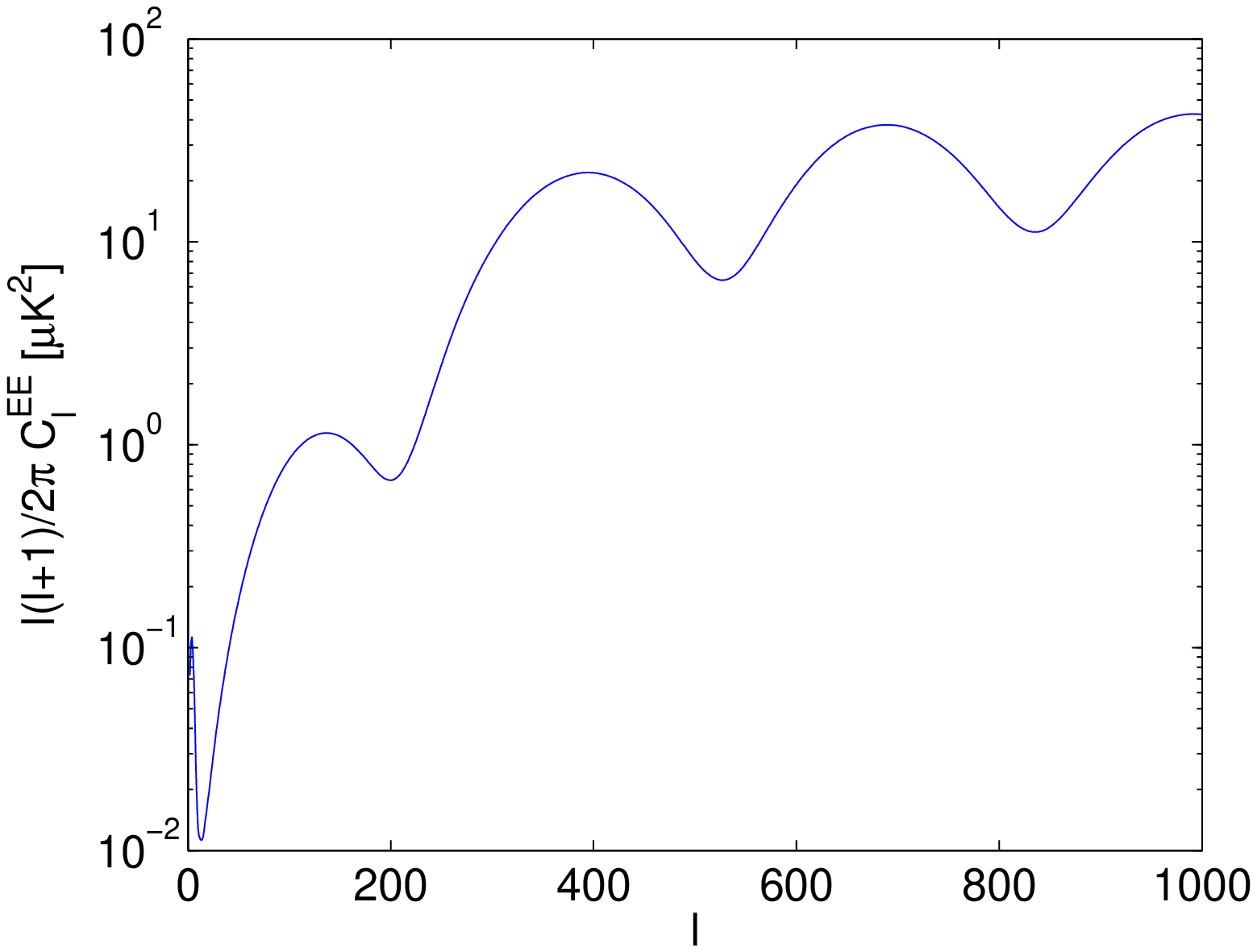}
\includegraphics[scale=.5]{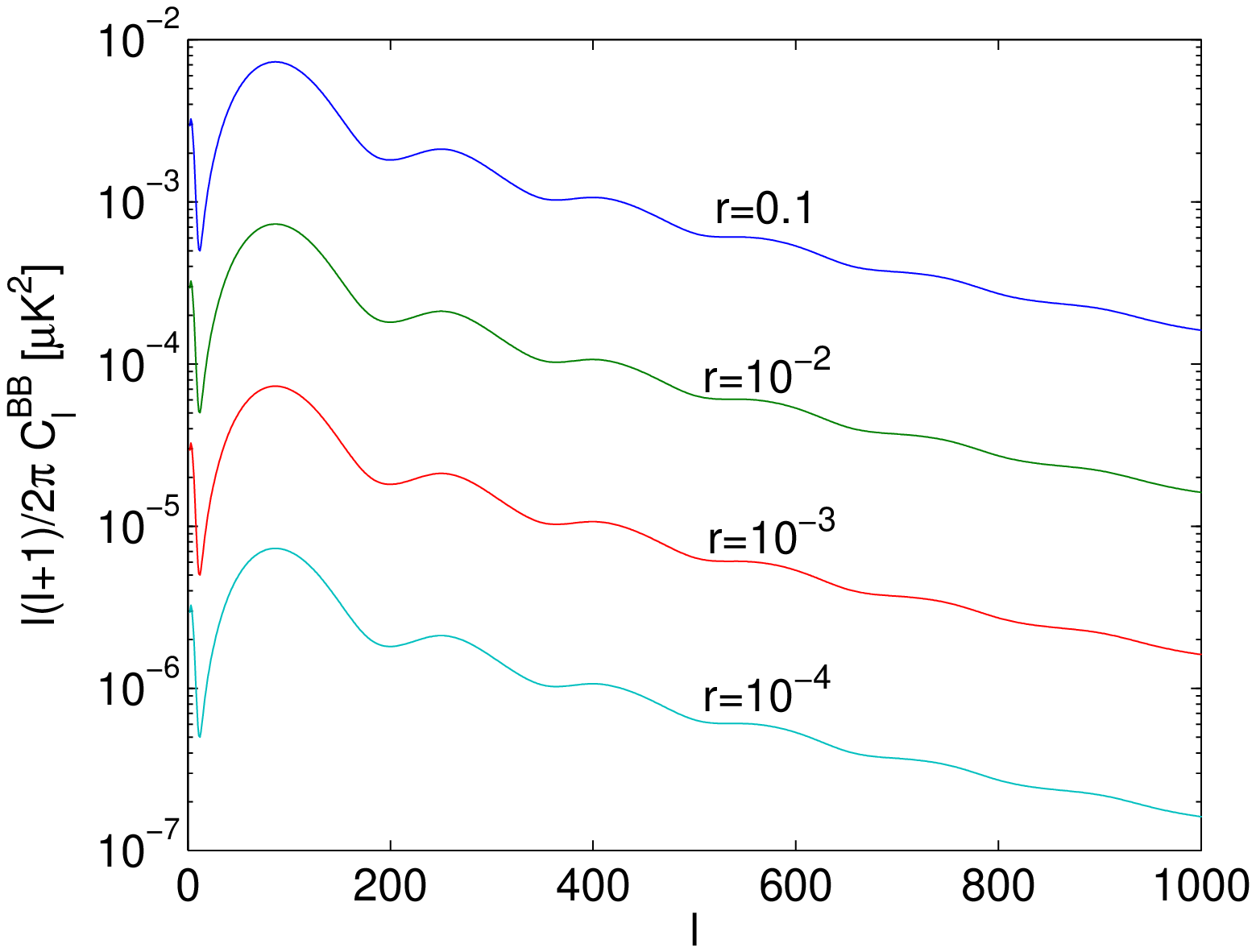}
\caption{the power spectrum of E (top) and B (bottom): no lensing, B mode power spectrum is plotted for various tensor-to-scalar ratio $r$.} 
\label{Cl}
\end{figure}
For a Gaussian seed fluctuation model, decomposition coefficients of E and B mode satisfy the following statistical properties:
\begin{eqnarray} 
\langle a^*_{E,lm} a_{E,l'm'} \rangle &=& C^{EE}_l\,\delta_{ll'}\delta_{mm'},\\
\langle a^*_{B,lm} a_{B,l'm'} \rangle &=& C^{BB}_l\,\delta_{ll'}\delta_{mm'},
\end{eqnarray}
where $\langle\ldots\rangle$ denotes an ensemble average.
In Fig. \ref{Cl}, we show unlensed $C^{EE}_l$ and  $C^{BB}_l$ of the WMAP concordance $\Lambda$CDM model for various tensor-to-scalar ratio $r$.

\section{E/B decomposition in pixel space}
\label{pixel_filter}
In this section, we are going to derive a pixel-space analogue of E/B decomposition. 
Using Eq. \ref{a2lm}, \ref{alm_E} and \ref{alm_B}, we may easily show Eq. \ref{E_map} and \ref{B_map} are equivalently given by:
\begin{eqnarray}
E(\mathbf {\hat n})&=&\sum  \sqrt{\frac{(l+2)!}{(l-2)!}} a_{E,lm}\,Y_{lm}(\mathbf {\hat n})\nonumber\\
&=&\sum  \sqrt{\frac{(l+2)!}{(l-2)!}} a^*_{E,lm}\,Y^*_{lm}(\mathbf {\hat n}),\nonumber\\
&=& -\frac{1}{2}\left(\int F_+(\mathbf {\hat n'},\mathbf{\hat n})\left[Q(\mathbf {\hat n'})-i U(\mathbf {\hat n'})\right] \,\mathrm d \Omega'\right.\nonumber\\
&+& \left.\int F_-(\mathbf {\hat n'},\mathbf{\hat n})\left[Q(\mathbf {\hat n'})+i U(\mathbf {\hat n'})\right] \,\mathrm d \Omega'\right), \label{E}
\end{eqnarray}
\begin{eqnarray}
B(\mathbf {\hat n})&=&\sum  \sqrt{\frac{(l+2)!}{(l-2)!}} a_{B,lm}\,Y_{lm}(\mathbf {\hat n}),\nonumber\\
&=&\sum  \sqrt{\frac{(l+2)!}{(l-2)!}} a^*_{B,lm}\,Y^*_{lm}(\mathbf {\hat n}),\nonumber\\
&=&\frac{i}{2}\left(\int F_+(\mathbf {\hat n'},\mathbf{\hat n})\left[Q(\mathbf {\hat n'})-i U(\mathbf {\hat n'})\right] \,\mathrm d \Omega'\right.\nonumber\\
&-& \left.\int F_-(\mathbf {\hat n'},\mathbf{\hat n})\left[Q(\mathbf {\hat n'})+i U(\mathbf {\hat n'})\right] \,\mathrm d \Omega'\right),\label{B}
\end{eqnarray}
where
\begin{eqnarray}
F_{\pm}(\mathbf {\hat n'},\mathbf{\hat n})&=&\sum_{lm}  \sqrt{\frac{(l+2)!}{(l-2)!}}\; {}_{\pm2}Y_{lm}(\mathbf {\hat n'})\;Y^*_{lm}(\mathbf {\hat n}). \label{F}
\end{eqnarray}
Therefore, we may identify $F_{\pm}(\mathbf {\hat n'},\mathbf{\hat n})$ as pixel-space filters for E/B decomposition.
Using the property of spin-$s$ spherical harmonics ${}_{s}Y^*_{lm}(\mathbf {\hat n})={}_{-s}Y_{l-m}(\mathbf {\hat n})$,
we may show the pair of the filter functions have complex conjugate relation:
\begin{eqnarray}
F_{+}(\mathbf {\hat n'},\mathbf{\hat n})=F^*_{-}(\mathbf {\hat n'},\mathbf{\hat n}). \label{conjugate}
\end{eqnarray}
\begin{figure}
\includegraphics[scale=.19]{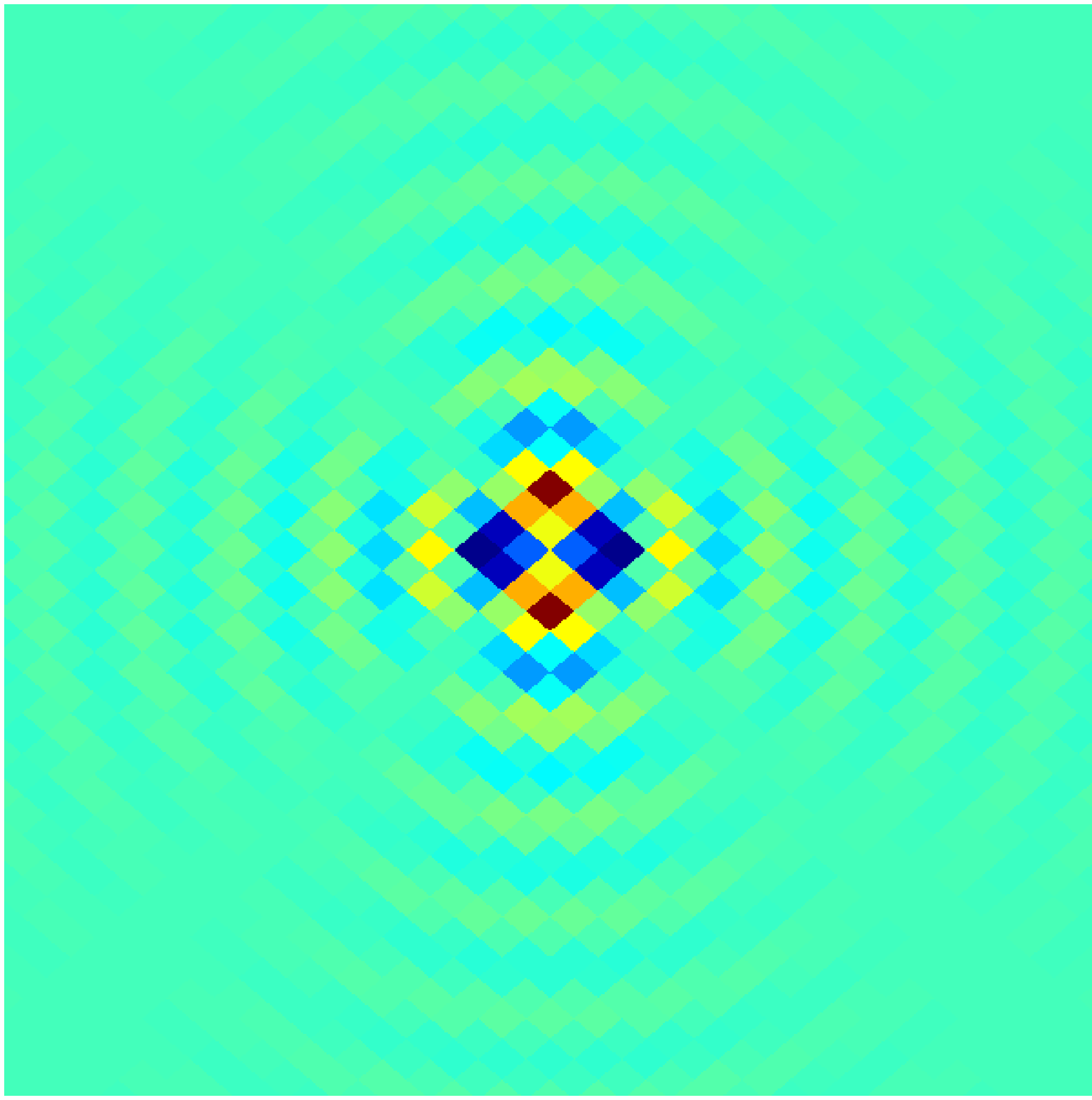}
\includegraphics[scale=.19]{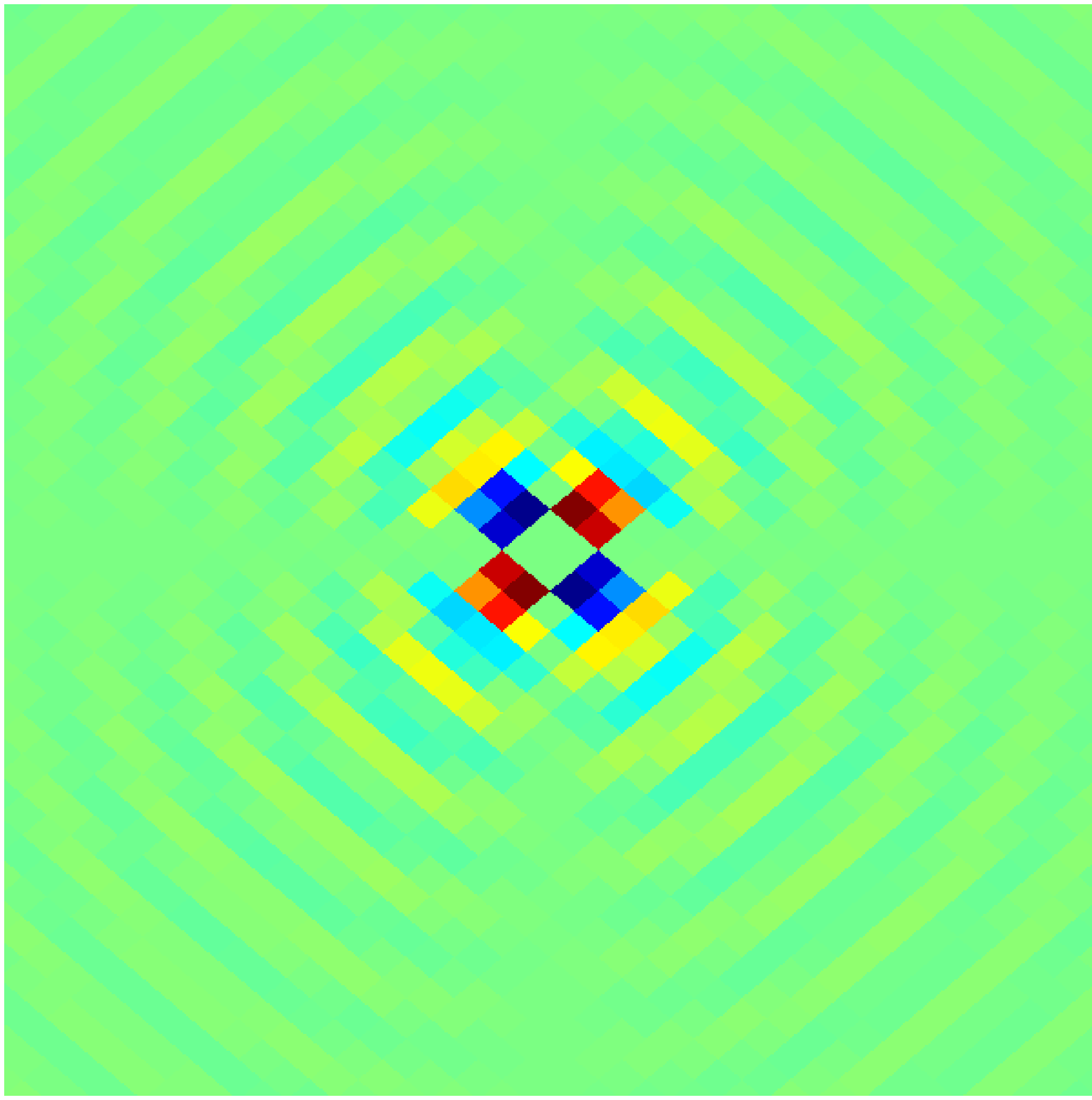}
\caption{Filter function: $\mathrm{Re}[F_{+}(\mathbf {\hat n'},\mathbf{\hat n})]$ (left), $\mathrm{Im}[F_{+}(\mathbf {\hat n'},\mathbf{\hat n})]$ (right) for a fixed $\mathbf n$, and $\mathbf n'$ spanning $2^\circ\times 2^\circ$ area, $F_{+}(\mathbf {\hat n'},\mathbf{\hat n})=\sum^{l\le 1024}_{lm} \sqrt{\frac{(l+2)!}{(l-2)!}}\;{}_{2}Y_{lm}(\mathbf {\hat n'})\;Y^*_{lm}(\mathbf {\hat n})$}
\label{RI}
\end{figure}
\begin{figure}
\centering\includegraphics[scale=.48]{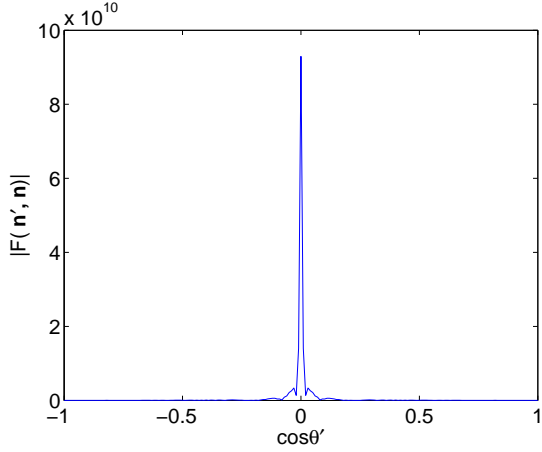}
\caption{Filter function: modulus $|F_{\pm}(\mathbf {\hat n'},\mathbf{\hat n})|$ for $(\theta=\pi/2,\phi=0)$, highly peaked at $\cos\theta'=0$}
\label{abs_F}
\end{figure}
In Fig. \ref{RI}, we show the real and imaginary part of $F_{+}(\mathbf {\hat n'},\mathbf{\hat n})$ for a fixed $\mathbf n$.
In Fig. \ref{abs_F}, we show one dimensional plot of $|F_{\pm}(\mathbf {\hat n'},\mathbf{\hat n})|$ for a fixed $\mathbf n$.
As shown in Fig. \ref{RI} and \ref{abs_F}, our pixel-space filter possesses sharp peaks around $\mathbf{\hat n}$.
Note that Eq. \ref{F} would approach $\delta(\mathbf {\hat n'}-\mathbf {\hat n})$, if $\sqrt{\frac{(l+2)!}{(l-2)!}}\; {}_{\pm2}Y_{lm}(\mathbf {\hat n'})$ were $Y_{lm}(\mathbf {\hat n'})$.
Using Eq. \ref{conjugate}, we may easily show Eq. \ref{E} and \ref{B} are equivalently given by:
\begin{eqnarray}
E(\mathbf {\hat n})&=& - \int \mathrm d \Omega'\,\mathrm{Re}\left[F_+(\mathbf {\hat n'},\mathbf{\hat n}) \left(Q(\mathbf {\hat n'})-i U(\mathbf {\hat n'})\right) \right],\nonumber\\
&=& - \int \mathrm d \Omega'\,\mathrm{Re}\left[F_-(\mathbf {\hat n'},\mathbf{\hat n}) \left(Q(\mathbf {\hat n'})+i U(\mathbf {\hat n'})\right) \right].\label{R_E} \nonumber
\end{eqnarray}

\begin{eqnarray}
B(\mathbf {\hat n})&=& -\int \mathrm d \Omega' \mathrm{Im}\left[F_+(\mathbf {\hat n'},\mathbf{\hat n}) \left(Q(\mathbf {\hat n'})-i U(\mathbf {\hat n'})\right) \right],\nonumber\\
&=&\int \mathrm d \Omega' \mathrm{Im}\left[F_-(\mathbf {\hat n'},\mathbf{\hat n})\left(Q(\mathbf {\hat n'})+i U(\mathbf {\hat n'})\right) \right].\label{R_B}\nonumber
\end{eqnarray}

\section{Incomplete sky coverage}
\label{cutsky}
Due to heavy foreground contamination, CMB polarization signal is not estimated reliably over a whole sky.
For instance, the WMAP team have subtracted diffuse foregrounds by template-fitting, and masked the regions that cannot be cleaned reliably.
In Fig. \ref{mask}, we show a foreground mask, which combines the WMAP team's polarization mask with the point source mask and shall be used for our simulation.
The E/B decomposition coefficients from a masked sky are given by:
\begin{eqnarray}
\tilde a_{E,lm}&=&-(\tilde a_{2,lm}+ \tilde a_{-2,lm})/2,\label{ialm_E}\\
\tilde a_{B,lm}&=&i(\tilde a_{2,lm} - \tilde a_{-2,lm})/2,\label{ialm_B}
\end{eqnarray}
where
\begin{eqnarray}
\tilde a_{\pm2,lm}=\int W(\mathbf {\hat n'})\left[Q(\mathbf {\hat n})\pm i U(\mathbf {\hat n})\right]\,{}_{\pm2}Y^*_{lm}(\mathbf {\hat n})\,\mathrm d \Omega.
\end{eqnarray}
and $W(\mathbf {\hat n'})$ is a foreground mask. Therefore, E/B maps reconstructed from incomplete sky are given by:
\begin{eqnarray}
\tilde E(\mathbf {\hat n})&=&\sqrt{\frac{(l+2)!}{(l-2)!}}\:\tilde a_{E,lm}\,Y_{lm}(\mathbf {\hat n}),\label{E_masked}\\
&=&-\frac{1}{2}\left(\int W(\mathbf {\hat n'})\,F_+(\mathbf {\hat n'},\mathbf{\hat n})\left[Q(\mathbf {\hat n'})-i U(\mathbf {\hat n'})\right] \,\mathrm d \Omega'\right.\nonumber\\
&+& \left.\int  W(\mathbf {\hat n'})\,F_-(\mathbf {\hat n'},\mathbf{\hat n})\left[Q(\mathbf {\hat n'})+i U(\mathbf {\hat n'})\right] \,\mathrm d \Omega'\right), \nonumber
\end{eqnarray}
\begin{eqnarray}
\tilde B(\mathbf {\hat n})&=&\sqrt{\frac{(l+2)!}{(l-2)!}}\:\tilde a_{B,lm}\,Y_{lm}(\mathbf {\hat n}),\label{B_masked}\\ 
&=&\frac{i}{2}\left(\int  W(\mathbf {\hat n'})\,F_+(\mathbf {\hat n'},\mathbf{\hat n})\left[Q(\mathbf {\hat n'})-i U(\mathbf {\hat n'})\right] \,\mathrm d \Omega'\right.\nonumber\\
&-& \left.\int  W(\mathbf {\hat n'})\,F_-(\mathbf {\hat n'},\mathbf{\hat n})\left[Q(\mathbf {\hat n'})+i U(\mathbf {\hat n'})\right] \,\mathrm d \Omega'\right).\nonumber
\end{eqnarray}
Since filter functions $F_{\pm}(\mathbf {\hat n'},\mathbf{\hat n})$ are sharply peaked around $\mathbf {\hat n}$, certain pixels far away from masked regions may contain negligible E/B mixing, and vice versa. Equivalently, E/B mixing is localized in pixels close to the masked regions.
For higher pixel resolution, Eq. \ref{F} contains summation up to higher $l$, which makes the peak of the filter function sharper.
Therefore, E/B mixing decreases with increase in pixel resolution. 
The spherical harmonic method (the first line of Eq. \ref{E_masked} and \ref{B_masked}) are much faster than the pixel-space method  (the second line), while they are mathematically equivalent. 
Therefore, we are going to rely on spherical harmonic transformation method for our simulation in the next section.
However, it should be kept in mind that the pixel-space approach have provided useful insights on E/B decomposition of incomplete sky.

\begin{figure}
\centering\includegraphics[scale=.5]{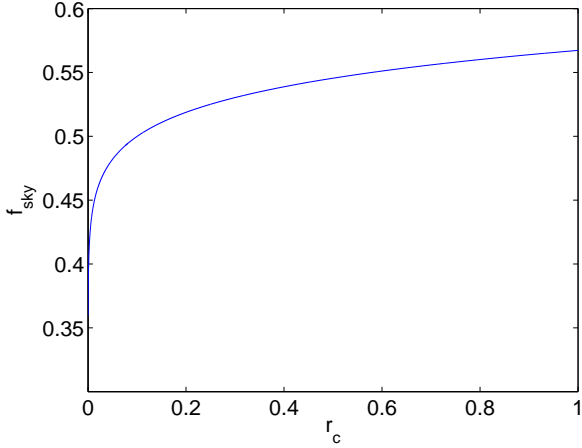}
\caption{effective sky fraction $f_{\mathrm{sky}}$ for various cut level $r_c$}
\label{f_sky}
\end{figure}

\begin{figure}
\centering\includegraphics[scale=.5]{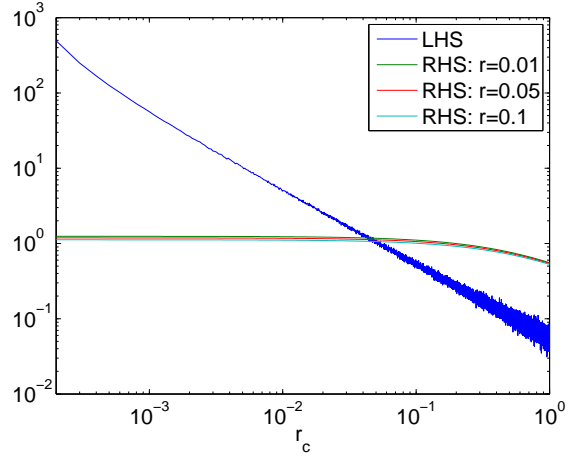}
\caption{Numerical solution of Eq. \ref{derivative} for various $r$ and the noise level of Planck HFI instrument: two plots represent the Left Hand Side (LHS) and Right Hand Side (RHS) of Eq. \ref{derivative}.}
\label{dlnf}
\end{figure}

Using Eq. \ref{B_masked}, we may show the expected power of $\tilde B(\mathbf {\hat n})$ is given by:
\begin{eqnarray}
\lefteqn{\langle \tilde B^2(\mathbf {\hat n})\rangle=\frac{1}{4}\int \mathrm d \Omega' \mathrm d \Omega''\:W(\mathbf {\hat n'})W(\mathbf {\hat n''})\times} \label{B_power}\\
&&(F_+(\mathbf {\hat n'},\mathbf{\hat n})F_-(\mathbf {\hat n''},\mathbf{\hat n})\langle (Q(\mathbf {\hat n'})-i U(\mathbf {\hat n'})) (Q(\mathbf {\hat n''})+i U(\mathbf {\hat n''}))\rangle\nonumber\\
&&+F_-(\mathbf {\hat n'},\mathbf{\hat n})F_+(\mathbf {\hat n''},\mathbf{\hat n})\langle (Q(\mathbf {\hat n'})+i U(\mathbf {\hat n'})) (Q(\mathbf {\hat n''})-i U(\mathbf {\hat n''}))\rangle\nonumber\\
&&-F_+(\mathbf {\hat n'},\mathbf{\hat n})F_+(\mathbf {\hat n''},\mathbf{\hat n})\langle (Q(\mathbf {\hat n'})-i U(\mathbf {\hat n'})) (Q(\mathbf {\hat n''})-i U(\mathbf {\hat n''}))\rangle \,\nonumber\\
&&-F_-(\mathbf {\hat n'},\mathbf{\hat n})F_-(\mathbf {\hat n''},\mathbf{\hat n})\langle (Q(\mathbf {\hat n'})+i U(\mathbf {\hat n'})) (Q(\mathbf {\hat n''})+i U(\mathbf {\hat n''}))\rangle)\nonumber.
\end{eqnarray}
where $\langle \ldots \rangle$ denotes an ensemble average, and
\begin{eqnarray}
\lefteqn{\langle (Q(\mathbf {\hat n'})+ i U(\mathbf {\hat n'})) (Q(\mathbf {\hat n''})+ i U(\mathbf {\hat n''}))\rangle} \label{QU_correlation1}\\
&=&\sum_l \sqrt{\frac{2l+1}{4\pi}} (C^{EE}_l+C^{BB}_l)\;{}_2Y_{l,-2}(\beta,0)\,e^{2i(\alpha+\gamma)},\nonumber
\end{eqnarray}
\begin{eqnarray}
\lefteqn{\langle (Q(\mathbf {\hat n'})\mp i U(\mathbf {\hat n'})) (Q(\mathbf {\hat n''})\pm i U(\mathbf {\hat n''}))\rangle} \label{QU_correlation2}\\
&=&\sum_l \sqrt{\frac{2l+1}{4\pi}} (C^{EE}_l+C^{BB}_l)\;{}_2Y_{l,-2}(\beta,0)\,e^{\pm 2i(\alpha-\gamma)},\nonumber
\end{eqnarray}
and $\beta$ is the separation angle between $\hat n'$ and $\hat n''$, 
$\alpha$ and $\gamma$ are the rotation angles respectively, which align $\hat e_{\theta}$ at $\hat n'$ and $\hat n''$with the great circle passing through $\hat n'$ and $\hat n''$ (refer to Fig. 1 of \cite{Ng_correlation_pol} for a geometrical diagram). 
Taking into account Eq. \ref{B_power}, \ref{QU_correlation1} and \ref{QU_correlation2}, we may easily show that the expected leakage power at $\mathbf {\hat n}$ is given by
\begin{eqnarray}
\lefteqn{\langle \tilde B^2_E(\mathbf {\hat n})\rangle=\frac{1}{4}\int \mathrm d \Omega' \mathrm d \Omega''\:W(\mathbf {\hat n'})W(\mathbf {\hat n''})\times} \label{leakage_power}\\
&&(F_+(\mathbf {\hat n'},\mathbf{\hat n})F_-(\mathbf {\hat n''},\mathbf{\hat n})\langle (Q_E(\mathbf {\hat n'})-i U_E(\mathbf {\hat n'})) (Q_E(\mathbf {\hat n''})+i U_E(\mathbf {\hat n''}))\rangle\nonumber\\
&&+F_-(\mathbf {\hat n'},\mathbf{\hat n})F_+(\mathbf {\hat n''},\mathbf{\hat n})\langle (Q_E(\mathbf {\hat n'})+i U_E(\mathbf {\hat n'})) (Q_E(\mathbf {\hat n''})-i U_E(\mathbf {\hat n''}))\rangle\nonumber\\
&&-F_+(\mathbf {\hat n'},\mathbf{\hat n})F_+(\mathbf {\hat n''},\mathbf{\hat n})\langle (Q_E(\mathbf {\hat n'})-i U_E(\mathbf {\hat n'})) (Q_E(\mathbf {\hat n''})-i U_E(\mathbf {\hat n''}))\rangle \,\nonumber\\
&&-F_-(\mathbf {\hat n'},\mathbf{\hat n})F_-(\mathbf {\hat n''},\mathbf{\hat n})\langle (Q_E(\mathbf {\hat n'})+i U_E(\mathbf {\hat n'})) (Q_E(\mathbf {\hat n''})+i U_E(\mathbf {\hat n''}))\rangle)\nonumber.
\end{eqnarray}
where 
\begin{eqnarray}
\lefteqn{\langle (Q_E(\mathbf {\hat n'})+ i U_E(\mathbf {\hat n'})) (Q_E(\mathbf {\hat n''})+ i U_E(\mathbf {\hat n''}))\rangle} \label{QU_E1}\\
&=&\sum_l \sqrt{\frac{2l+1}{4\pi}} C^{EE}_l\;{}_2Y_{l,-2}(\beta,0)\,e^{2i(\alpha+\gamma)},\nonumber
\end{eqnarray}
\begin{eqnarray}
\lefteqn{\langle (Q_E(\mathbf {\hat n'})\mp i U_E(\mathbf {\hat n'})) (Q_E(\mathbf {\hat n''})\pm i U_E(\mathbf {\hat n''}))\rangle} \label{QU_E2}\\
&=&\sum_l \sqrt{\frac{2l+1}{4\pi}} C^{EE}_l\;{}_2Y_{l,-2}(\beta,0)\,e^{\pm 2i(\alpha-\gamma)}.\nonumber
\end{eqnarray}
Therefore, we may diagnose ambiguous pixels (i.e heavy E/B mixing) by comparing $\langle \tilde B^2_E(\mathbf {\hat n})\rangle$ with $\langle \tilde B^2_B(\mathbf {\hat n})\rangle$,
where $\langle \tilde B^2_B(\mathbf {\hat n})\rangle$ is the local power contributed by B mode and given by replacing `E' with `B' in Eq. \ref{leakage_power}, \ref{QU_E1} and \ref{QU_E2}.
However, estimating Eq. \ref{leakage_power}, \ref{QU_E1} and \ref{QU_E2} is prohibitively complicated.
Therefore, we are going to resort to Monte-Carlo simulations in order to estimate $\langle \tilde B^2_E(\mathbf {\hat n})\rangle/\langle \tilde B^2_B(\mathbf {\hat n})\rangle$.
Depending $\langle \tilde B^2_E(\mathbf {\hat n})\rangle/\langle \tilde B^2_B(\mathbf {\hat n})\rangle$, we may classify the pixel at $\mathbf{\hat n}$ as `pure' and `ambiguous'.
To be specific, we may retain pixels satisfying:
\begin{eqnarray}
\frac{\langle \tilde B^2_E(\mathbf {\hat n})\rangle}{\langle \tilde B^2_B(\mathbf {\hat n})\rangle}<\frac{r_c}{r},
\end{eqnarray}
where $r$ is the assumed tensor-to-scalar ratio of Monte-Carlo simulation, from which $\langle \tilde B^2_E(\mathbf {\hat n})\rangle/\langle \tilde B^2_B(\mathbf {\hat n})\rangle$ is estimated. Therefore, the level of leakage in retained pixels is comparable to the primordial B mode power spectrum of tensor-to-scalar ratio $r_c$.
In Fig. \ref{f_sky}, we show the sky fraction for various $r_c$, given a foreground mask shown in Fig. \ref{mask}. 
Since sky fraction decreases with lower $r_c$, we may not simply set $r_c$ to a lowest value.
Therefore, we need to derive an optimal $r_c$, which minimizes the estimation error. 
The estimation error of B mode power spectrum is given by:
\begin{eqnarray}
\Delta C^{BB}_l&=&\frac{2}{(2l+1)f_{\mathrm{sky}}}(C^{BB}_l+ \tilde C^{EE}_l +N_l), \nonumber\\
&\approx& \frac{2}{(2l+1)f_{\mathrm{sky}}}(C^{BB}_l+ \frac{r_c}{r}\,C^{BB}_l +N_l),
\end{eqnarray}
where $N_l$ is noise power spectrum. 
Note that the leakage does not bias the B mode power spectrum estimation, but increases the variance, when the power spectrum estimation is made by a pseudo-$C_l$ method and leakage is taken care of \citep{Master_power,Grain_mixing}. By requiring $\partial\,\Delta C^{BB}_l/\partial r_c=0$, we get
\begin{eqnarray}
\frac{\partial \ln f_{\mathrm{sky}}}{\partial r_c}=\frac{1}{r+ r_c +N_l/C^{BB}_l(r=1)} \label{derivative}.
\end{eqnarray}
In Fig. \ref{dlnf}, we plot the left and right hand side of Eq. \ref{derivative} for the noise level of Planck HFI instrument, and the multipole $l=86$, which is the peak multipole of primordial B mode power spectrum.  From Fig. \ref{dlnf}, we find curves intersect at $r_c\approx 4\times 10^{-2}$ with weak dependence on $r$.
It should be noted that the weak dependence is due to the low signal-to-noise ratio of the considered experiment (i.e. $N_l/C^{BB}_l(r=1)\gg0$), and the dependence on $r$ is not weak in general. We are going to use $r_c\approx 4\times 10^{-2}$ for the simulation in the next section.

\section{Application to simulated data}
\label{simulation}
\begin{figure}
\centering\includegraphics[scale=.27]{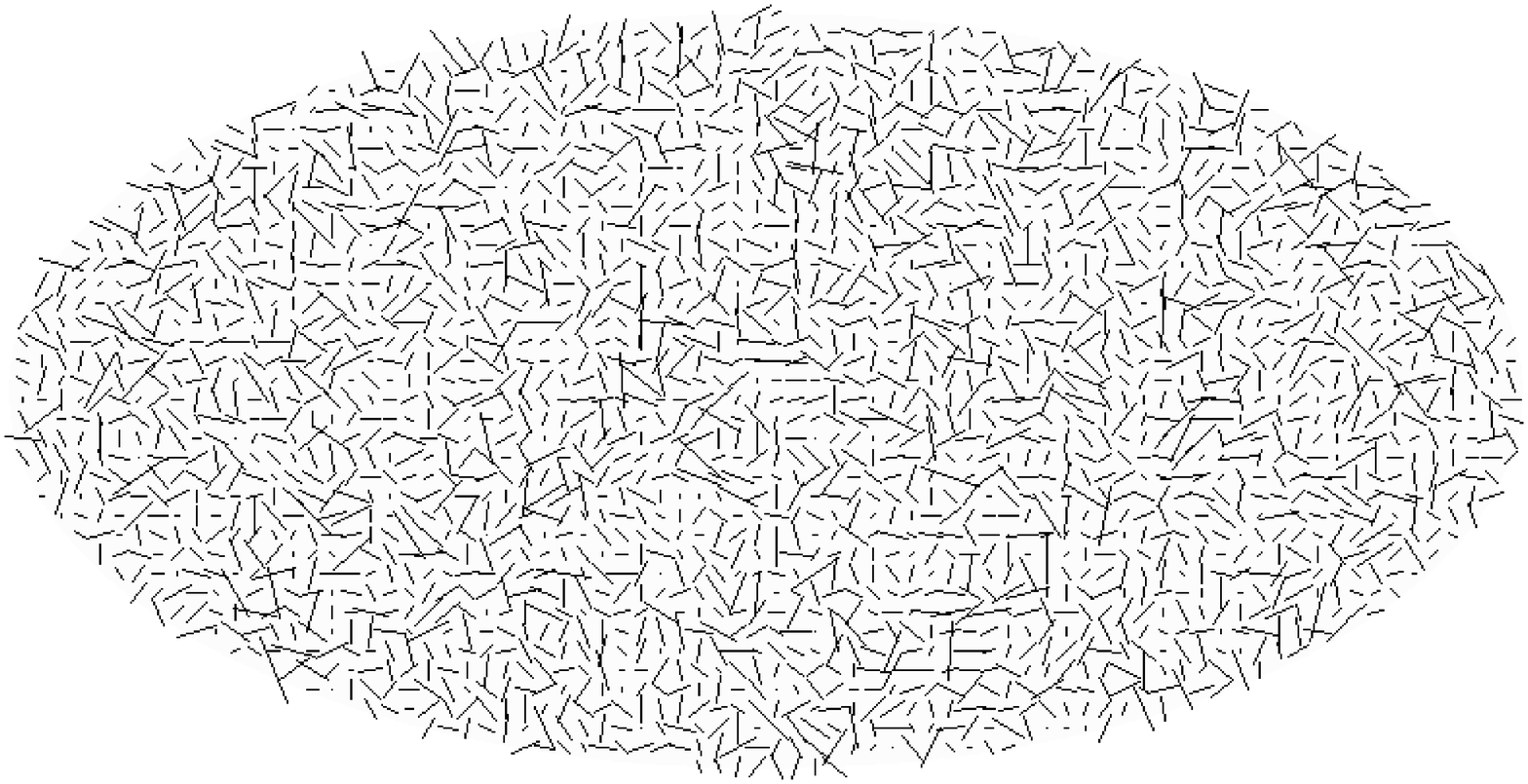}
\caption{Input polarization map: E mode polarization only}
\label{input}
\end{figure}

Using the WMAP concordance $\Lambda$CDM model, we have simulated Stokes parameter Q and U over a whole-sky with a HEALPix pixel resolution (Nside=1024) and $10'$ FWHM beam. 
We have made the inputmap to contain no B mode polarization. Therefore, any non-zero values in output B map are attributed to leakage.
We show our simulated polarization map in Fig. \ref{input}, where the orientation and length of headless arrows indicates polarization angle and amplitude respectively. 
Note that the polarization map shows only gradient-like patterns, because they contain only E mode polarization.
\begin{figure}
\centering\includegraphics[scale=.27]{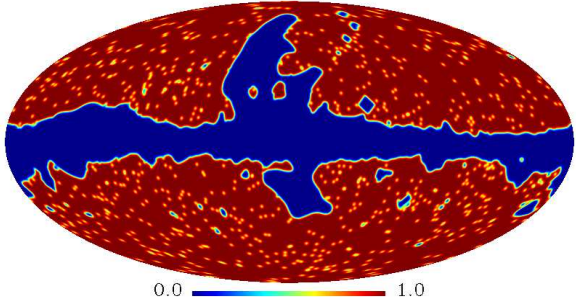}
\caption{foreground mask: smoothed with $1.5^\circ$ FWHM Gaussian kernel.}
\label{mask}
\end{figure}
It is well-known that E/B mixing increases with the length of cut sky boundary \citep{Bunn:EB-Separation}. 
We have combined the WMAP team's polarization mask with the point source mask, and prograded it to Nside=1024.
In order to reduce sharp boundaries, we have smoothed the mask with $1.5^\circ$ FWHM Gaussian kernel.
We have referred to the WMAP team's boundary smoothing process of Internal Linear Combination map \citep{WMAP3:temperature}.
Nevertheless, it should be noted that smoothed boundary is not essential to our method, and further improvement may be possible by using more sophisticated smoothing kernel \citep{edge_taper}. In Fig. \ref{mask}, we show our smoothed mask, whose sky fraction amounts to $0.71$.

\begin{figure}
\centering\includegraphics[scale=.27]{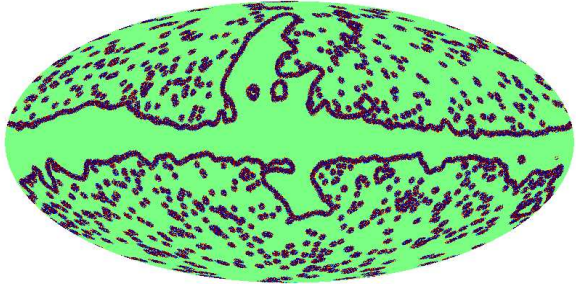}
\caption{output $\tilde B$ map from incomplete sky}
\label{output1}
\end{figure}
In Fig. \ref{output1}, we show a $\tilde B$ map, which we have produced from the masked polarization map.
Using $\langle \tilde B^2_E(\mathbf {\hat n})\rangle/\langle \tilde B^2_B(\mathbf {\hat n})\rangle$ estimated from $10^3$ simulation, we have diagnosed ambiguous pixels, and retained pixels of $\langle \tilde B^2_E(\mathbf {\hat n})\rangle/\langle \tilde B^2_B(\mathbf {\hat n})\rangle <4\times 10^{-2}/r$.
\begin{figure}
\centering\includegraphics[scale=.27]{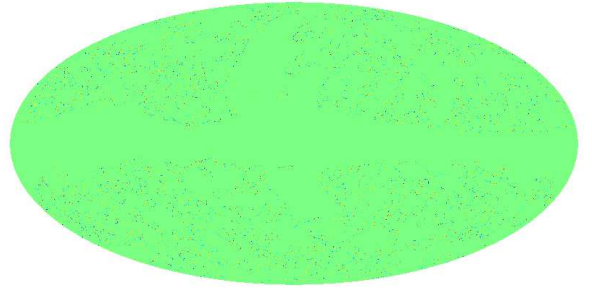}
\caption{filerted $\tilde B$ map: masked by a foreground mask, and pixels of $\langle \tilde B^2_{E}(\mathbf {\hat n})\rangle/\langle \tilde B^2_{B}(\mathbf {\hat n})\rangle>4\times 10^{-2}$ are set to zero.}
\label{output2}
\end{figure}
\begin{figure}
\centering\includegraphics[scale=.54]{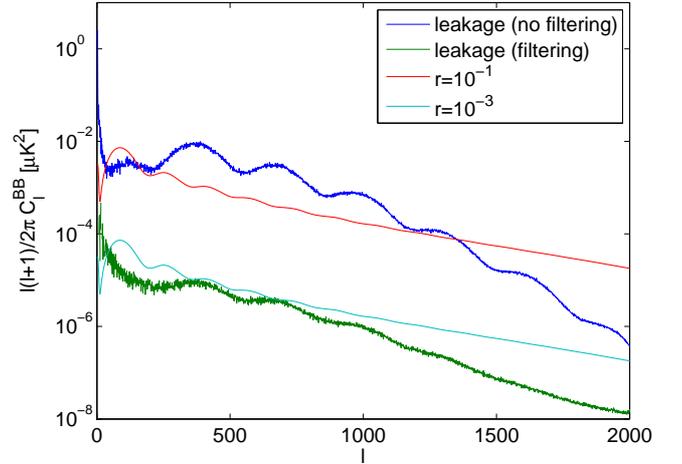}
\caption{leakage power spectrum and primordial B mode power of various tensor-to-scalar ratio $r$: 
a blue curve correspond to the leakage power estimated without ambiguous pixel filtering (Fig. \ref{output1}), a green curve to the leakage power estimated with ambiguous pixel filtering (Fig. \ref{output2}).}
\label{leakage}
\end{figure}
In Fig. \ref{output2}, we show the B map, where ambiguous pixels are excluded.
We find retained pixels of Fig. \ref{output2} amount to sky fraction $f_{\mathrm{sky}}=0.48$.
From the retained pixels, we have estimated the leakage power spectrum by pseudo $C_l$ method \citep{pseudo_Cl,MASTER}.
Power spectrum is usually estimated by a pseudo $C_l$ method at high multipoles ($l> 30$), while by maximum likelihood method or Gibbs sampling at low multipoles ($l\le 30$) \citep{Bond:likelihood,Gibbs_power,WMAP3:temperature,hybrid_estimation}.
However, we find pseudo $C_l$ method at low multipoles is good enough for our need, since we do not intend accurate estimation of likelihood function.
Besides that, we are mainly interested in leakage power at ($l> 30$), because primordial B mode power spectrum has a peak around multipoles $l\sim 90$ (see Fig. \ref{Cl}).
In Fig. \ref{leakage}, we show the leakage power spectrum and B mode power spectrum of various tensor-to-scalar ratio $r$. Fig. \ref{leakage} shows leakage power (green) is smaller than B mode power spectrum of $r=10^{-3}$ at wide range of multipoles ($50\la l \la 2000$), when ambiguous pixels are excluded (i.e. the B map in Fig. \ref{output2}).
In Fig. \ref{leakage}, we also show the leakage power (blue) estimated without filtering (i.e. the B map in Fig. \ref{output1}).
Obviously, we have reduced leakage significantly by excluding ambiguous pixels.

\section{scale-dependence of leakage}
\label{scale}
In order to reduce leakage, we have removed ambiguous pixels in a scale-independent way.
On the other hand, it is known that leakage has some dependence on scales as well as real-space.
Specifically, leakage of low $l$ extends over large area, while leakage of high $l$ is often confined to the small area nearby the boundary.
Therefore, one may argue that our method does not reduce the leakage at low $l$ as effectively as that of high $l$ or leads to unnecessary loss of information.
However, as shown in Fig. \ref{output1}, leakage is relatively localized in pixel space, and we were able to reduced leakage very significantly at wide range of multipoles, while retaining sky fraction $0.48$.
Besides that, our simulation shows the leakage at lowest $l$ is reduced significantly as well.
It is also possible to implement a further leakage reduction in a scale-dependent way, after ambiguous pixels are removed.
Nevertheless, a hybrid method, which exploits scale and position dependence simultaneously, may be most optimal.
A wavelet approach may be promising for such implementation, since wavelet functions are, in general, well-localized in harmonic space as well as pixel space. We defer a rigorous investigation to a separate publication.

\section{Discussion}
\label{Discussion}
We have investigated E/B decomposition in pixel space, and shown that we may produce E/B decomposed maps by convolving polarization maps with certain filter functions of a sharp peak. We find that E/B mixing due to incomplete sky is localized in pixel-space, and negligible in the regions far away from masked area. 
By estimating the expected local leakage power and comparing it with the expected pure mode power, we have diagnosed ambiguous pixels and excluded them.
Our criteria for ambiguous pixels (i.e. $r_c$) is associated with the tensor-to-scalar ratio of B mode power spectrum, which the leakage power is comparable to.
The estimation error $\Delta C_l$ may increases with lower $r_c$, because sky fraction decreases.
Therefore, we have solved  $\partial \Delta C_l/\partial r_c=0$ and obtained the optimal $r_c$, which minimizes the estimation error, given a foreground mask and noise level.
We have applied our method to simulated maps blocked by a foreground mask. 
Simulation shows that leakage power is subdominant in comparison with unlensed B mode power spectrum of  $r\sim 1\times10^{-3}$
at wide range of multipoles ($50\la l \la 2000$), while pixels of sky fraction $0.48$ are retained. 
We may apply our method equally to small sky patch observation, by treating unobserved sky as masked region. 
From simulation with sky patch of simple symmetric shape, we have confirmed our method reduce E/B mixing very effectively.
A rigorous investigation is deferred to a separate publication.

Noise is slightly correlated from pixel to pixel in E and B maps, even when interpixel correlation is absent in Q and U maps.
However, this interpixel noise correlation induced by E/B decomposition is not confined to our method, but E/B decomposition in general.
Besides that, we find interpixel noise correlation may be neglected without sacrificing the accuracy of error analysis (refer to Appendix \ref{noise} for details).
Therefore, it does not limit the applicability of our method. 

Current observations such as WMAP were unable to detect B mode polarization. 
Therefore, we did not attempt to apply our method to observation data. 
When Planck polarization data of high Signal-to-Noise-Ratio (SNR) are available in near future, we may apply our method to the data, and be able to detect B mode polarization.

\begin{acknowledgements}
We are grateful to the anonymous referee for thorough reading and helpful comments, which leads to significant improvement of this work.
We acknowledge the use of the Legacy Archive for Microwave Background Data Analysis (LAMBDA) and the HEALPix package \citep{HEALPix:Primer,HEALPix:framework}.
This work is supported by FNU grant 272-06-0417, 272-07-0528 and 21-04-0355. 
This work is supported in part by Danmarks Grundforskningsfond, which allowed the establishment of the Danish Discovery Center.
\end{acknowledgements}

\begin{appendix}
\section{error analysis}
\label{noise}
Power spectrum is usually estimated by pseudo-$C_l$ method at high multipoles \citep{pseudo_Cl,MASTER,hybrid_estimation,WMAP5:powerspectra,WMAP7:powerspectra}.
According to pseudo-$C_l$ method, we may estimate power spectrum as follows \citep{pseudo_Cl,MASTER}:
\begin{eqnarray}
\hat C^{BB}_l=\sum_{l'} (\mathbf M^{-1})_{l l'} \tilde C^{BB}_{l'}.
\end{eqnarray}
where
\begin{eqnarray}
\mathbf M_{l l'}=\frac{2l'+1}{4\pi} \sum_{l''m''} |w_{l''m''}|^2\left\{\left(\begin{array}{ccc}l&l'&l''\\2&0&-2\end{array}\right)-\left(\begin{array}{ccc}l&l'&l''\\-2&0&2\end{array}\right)\right\}^2,
\end{eqnarray}
and
\begin{eqnarray}
w_{l''m''}=\int \mathrm d \Omega\,W(\mathbf {\hat n})Y^*_{l''m''}(\mathbf {\hat n}),
\end{eqnarray}
with $W(\mathbf {\hat n})$ being a foreground mask function.
The pseudo-quantity $\tilde C_l$ and $\tilde a^j_{lm}$ are given by:
\begin{eqnarray}
\tilde C_l=\frac{1}{2l+1}\sum_m \tilde a^{j}_{lm} (\tilde a^{k}_{lm})^*,\label{pseudo_Cl}
\end{eqnarray}
\begin{eqnarray}
\tilde a^j_{lm}=\int \mathrm d \Omega\,W(\mathbf {\hat n})\,\Delta^j(\mathbf {\hat n}) Y^*_{lm}(\mathbf {\hat n}),\label{pseudo_alm}
\end{eqnarray}
where $j$ and $k$ refers to a DA-year combination and $\Delta(\mathbf {\hat n})$ refers to data (T, E or B).
We may split  $\tilde a^{j}_{lm}$ into signal and noise: 
\begin{eqnarray}
\tilde a^{j}_{lm}=\tilde a_{lm} +\tilde N^{j}_{lm}. \label{alm}
\end{eqnarray}
The noise part is given by:
\begin{eqnarray}
\tilde N_{lm}=\frac{4\pi}{n_{\mathrm{pix}}}\sum_{i} N_i\,Y^i_{lm}, \label{Nlm}
\end{eqnarray}
where a pixel index $i$ runs over pixels outside a foreground mask, and $N_i$ refers to noise at $i$th pixel.
By central limit theorem \citep{Arfken,Math_methods}, $\tilde N^{j}_{lm}$ follows a Gaussian distribution, and is uncorrelated among distinct $j$th DA-year data (i.e. $\langle \tilde N^{j}_{lm} (\tilde N^{k}_{lm})^*\rangle \propto \delta_{jk}$).

If cross power spectra are used (i.e. $j\ne k$), noise does not bias estimation, and its statistical properties need to be known only for errors analysis \citep{WMAP3:temperature}.
In order to understand the effect of noise on error analysis, let us consider covariance of $\hat C_l$:
\begin{eqnarray}
\mathrm{Cov}(\hat C_{l_1},\hat C_{l_2})=\langle \hat C_{l_1} \hat C_{l_2}\rangle - C_{l_1} C_{l_2},\label{Cov}
\end{eqnarray}
where
$C_l$ is a theoretical power spectrum and
\begin{eqnarray}
\langle \hat C_{l_1} \hat C_{l_2}\rangle&=&\langle \sum_{l} (\mathbf M^{-1})_{l_1 l}\:\tilde C_{l} \sum_{l'} (\mathbf M^{-1})_{l_2 l'}\:\tilde C_{l'} \rangle, \nonumber\\
&=& \sum_{l} (\mathbf M^{-1})_{l_1 l} \sum_{l'} (\mathbf M^{-1})_{l_2 l'} \langle\tilde C_{l} \tilde C_{l'} \rangle. \label{Cl1l2}
\end{eqnarray}
Using Eq. \ref{pseudo_Cl} and \ref{alm}, we find
\begin{eqnarray}
\lefteqn{\langle \tilde C_{l} \tilde C_{l'} \rangle = A }\label{Cll}\\
&+&\frac{1}{(2l+1)(2l'+1)}\sum_{mm'} \langle \tilde N^{j}_{lm} (\tilde N^{j}_{l'm'})^* \rangle \langle \tilde N^{k}_{lm} (\tilde N^{k}_{l'm'})^* \rangle,\nonumber
\end{eqnarray}
where $A$ denotes terms irrelevant to noise.
Therefore, we need to estimate noise covariance $\langle \tilde N^{j}_{lm} (\tilde N^{j}_{l'm'})^* \rangle$ in order to estimate Eq. \ref{Cll}.
First, let us consider diagonal elements of noise covariance:
\begin{eqnarray}
\lefteqn{\langle \tilde N_{lm} \tilde N^*_{lm} \rangle} \label{Nlm_diag}\\
&=&\frac{16\pi^2}{n^2_{\mathrm{pix}}}\sum_{i} \langle N^2_i\rangle |Y^i_{lm}|^2+\frac{32\pi^2}{n^2_{\mathrm{pix}}}\sum_{i}\sum_{i'>i} \langle  N_i N_{i'}\rangle Y^i_{lm}(Y^{i'}_{lm})^*.\nonumber
\end{eqnarray}
The main contribution of Eq. \ref{Nlm_diag} comes from the first term, because of cancellation through summation in the second term.
Therefore, we find with good approximation:
\begin{eqnarray*}
\langle \tilde N_{lm} \tilde N^*_{lm} \rangle\approx\frac{16\pi^2}{n^2_{\mathrm{pix}}}\sum_{i} \langle N^2_i\rangle |Y^i_{lm}|^2.
\end{eqnarray*}
Off-diagonal elements of noise covariance are given by:
\begin{eqnarray}
\langle \tilde N_{lm} \tilde N^*_{l'm'} \rangle&=&\frac{16\pi^2}{n^2_{\mathrm{pix}}}\sum_{i} \langle N^2_i \rangle Y^i_{lm}(Y^i_{l'm'})^*\label{Nlm_offdiag}\\
&+&\frac{32\pi^2}{n^2_{\mathrm{pix}}}\sum_{i}\sum_{i'>i} \langle  N_i N_{i'}\rangle Y^i_{lm}(Y^{i'}_{l'm'})^*.\nonumber
\end{eqnarray}
Comparing \ref{Nlm_offdiag} with Eq. \ref{Nlm_diag}, we may see the magnitude of off-diagonal elements is much smaller than that of diagonal elements, because cancellation through summation arise both in the first and the second term of Eq. \ref{Nlm_offdiag}.
Therefore, we find noise covariance as follow:
\begin{eqnarray}
\langle \tilde N_{lm} \tilde N^*_{l'm'} \rangle\approx \delta_{ll'} \delta_{mm'}\frac{16\pi^2}{n^2_{\mathrm{pix}}}\sum_{i} \langle N^2_i\rangle |Y^i_{lm}|^2.\label{Nlm_cov}
\end{eqnarray}
Using Eq. \ref{Cov}, \ref{Cl1l2}, \ref{Cll} and \ref{Nlm_cov}, we find covariance of $\hat C_l$: 
\begin{eqnarray}
\lefteqn{\mathrm{Cov}(\hat C_{l_1},\hat C_{l_2})\approx C +\sum_{l} (\mathbf M^{-1})_{l_1 l} (\mathbf M^{-1})_{l_2 l}} \label{Cov_approx}\\
&\times&\;\frac{256\pi^4}{n^4_{\mathrm{pix}} (2l+1)^2}\sum_m\sum_{ii'} \langle (N^j_i)^2\rangle \langle (N^k_{i'})^2\rangle |Y^i_{lm}|^2 |Y^{i'}_{lm}|^2,\nonumber
\end{eqnarray}
where $C$ denotes terms irrelevant to noise.
As shown in Eq. \ref{Cov_approx}, we may neglect interpixel noise correlation in computing covariance of $\hat C_l$, 
\end{appendix}

\bibliographystyle{aa}
\bibliography{/home/tac/jkim/Documents/bibliography}

\begin{thebibliography}{44}
\expandafter\ifx\csname natexlab\endcsname\relax\def\natexlab#1{#1}\fi

\bibitem[{Ade \& et~al.(2008)}]{QUaD1}
Ade, P. \& et~al. 2008, \apj, 674, 22

\bibitem[{Arfken \& Weber(2000)}]{Arfken}
Arfken, G.~B. \& Weber, H.~J. 2000, Mathematical Methods for Physicists, 5th
  edn. (San Diego, CA USA: Academic Press)

\bibitem[{Bond {et~al.}(1998)Bond, Jaffe, \& Knox}]{Bond:likelihood}
Bond, J.~R., Jaffe, A.~H., \& Knox, L. 1998, \prd, 57, 2117

\bibitem[{{Brown} \& et~al.(2009)}]{QUaD_improved}
{Brown}, M.~L. \& et~al. 2009, \apj, 705, 978

\bibitem[{Bunn {et~al.}(2003)Bunn, Zaldarriaga, Tegmark, \&
  de~Oliveira-Costa}]{Bunn:EB-Separation}
Bunn, E.~F., Zaldarriaga, M., Tegmark, M., \& de~Oliveira-Costa, A. 2003, 67,
  023501

\bibitem[{{Das} {et~al.}(2009){Das}, {Hajian}, \& {Spergel}}]{edge_taper}
{Das}, S., {Hajian}, A., \& {Spergel}, D.~N. 2009, \prd, 79, 083008

\bibitem[{Dodelson(2003)}]{Modern_Cosmology}
Dodelson, S. 2003, Modern Cosmology, 2nd edn. (Academic Press)

\bibitem[{Efstathiou(2006)}]{hybrid_estimation}
Efstathiou, G. 2006, \mnras, 370, 343

\bibitem[{{Eriksen} {et~al.}(2004){Eriksen}, {O'Dwyer}, {Jewell}, {Wandelt},
  {Larson}, {G{\'o}rski}, {Levin}, {Banday}, \& {Lilje}}]{Gibbs_power}
{Eriksen}, H.~K., {O'Dwyer}, I.~J., {Jewell}, J.~B., {et~al.} 2004, \apj, 155,
  227

\bibitem[{Gorski {et~al.}(2005)Gorski, Hivon, Banday, Wandelt, Hansen,
  Reinecke, \& Bartelman}]{HEALPix:framework}
Gorski, K.~M., Hivon, E., Banday, A.~J., {et~al.} 2005, \apj, 622, 759

\bibitem[{{Gorski} {et~al.}(1999){Gorski}, {Wandelt}, {Hansen}, {Hivon}, \&
  {Banday}}]{HEALPix:Primer}
{Gorski}, K.~M., {Wandelt}, B.~D., {Hansen}, F.~K., {Hivon}, E., \& {Banday},
  A.~J. 1999, ArXiv Astrophysics e-prints

\bibitem[{{Grain} {et~al.}(2009){Grain}, {Tristram}, \&
  {Stompor}}]{Grain_mixing}
{Grain}, J., {Tristram}, M., \& {Stompor}, R. 2009, \prd, 79, 123515

\bibitem[{Halverson {et~al.}(2002)Halverson, Leitch, Pryke, Kovac, Carlstrom,
  Holzapfel, Dragovan, Cartwright, Mason, Padin, Pearson, Shepherd, \&
  Readhead}]{DASI:II}
Halverson, N.~W., Leitch, E.~M., Pryke, C., {et~al.} 2002, \apj, 568, 38

\bibitem[{Hinderks \& et~al.(2008)}]{QUaD:instrument}
Hinderks, J. \& et~al. 2008, ArXiv e-prints

\bibitem[{Hinshaw \& et~al.(2007)}]{WMAP3:temperature}
Hinshaw, G. \& et~al. 2007, \apj, 170, 288

\bibitem[{{Hivon} {et~al.}(2002{\natexlab{a}}){Hivon}, {G{\'o}rski},
  {Netterfield}, {Crill}, {Prunet}, \& {Hansen}}]{Master_power}
{Hivon}, E., {G{\'o}rski}, K.~M., {Netterfield}, C.~B., {et~al.}
  2002{\natexlab{a}}, \apj, 567, 2

\bibitem[{{Hivon} {et~al.}(2002{\natexlab{b}}){Hivon}, {G{\'o}rski},
  {Netterfield}, {Crill}, {Prunet}, \& {Hansen}}]{MASTER}
{Hivon}, E., {G{\'o}rski}, K.~M., {Netterfield}, C.~B., {et~al.}
  2002{\natexlab{b}}, \apj, 567, 2

\bibitem[{K.~F. Riley M. P.~Hobson(2006)}]{Math_methods}
K.~F. Riley M. P.~Hobson, S. J.~B. 2006, Mathematical Methods for Physics and
  Engineering: A Comprehensive Guide, 3rd edn. (Cambridge University Press)

\bibitem[{Kamionkowski {et~al.}(1997)Kamionkowski, Kosowsky, \&
  Stebbins}]{Kamionkowski:Flm}
Kamionkowski, M., Kosowsky, A., \& Stebbins, A. 1997, \prd, 55, 7368

\bibitem[{{Kim}(2007{\natexlab{a}})}]{Kim:measuring_a2lm}
{Kim}, J. 2007{\natexlab{a}}, \mnras, 375, 625

\bibitem[{{Kim}(2007{\natexlab{b}})}]{Kim:optimization}
{Kim}, J. 2007{\natexlab{b}}, \mnras, 375, 615

\bibitem[{{Komatsu} {et~al.}(2010){Komatsu}, {Smith}, {Dunkley}, {Bennett},
  {Gold}, {Hinshaw}, {Jarosik}, {Larson}, {Nolta}, {Page}, {Spergel},
  {Halpern}, {Hill}, {Kogut}, {Limon}, {Meyer}, {Odegard}, {Tucker}, {Weiland},
  {Wollack}, \& {Wright}}]{WMAP7:Cosmology}
{Komatsu}, E., {Smith}, K.~M., {Dunkley}, J., {et~al.} 2010, ArXiv e-prints

\bibitem[{Kovac {et~al.}(2002)Kovac, Leitch, Pryke, Carlstrom, Halverson, \&
  Holzapfel}]{DASI:data}
Kovac, J., Leitch, E.~M., Pryke, C., {et~al.} 2002, Nature, 420, 772

\bibitem[{Kraus(1986)}]{Kraus:Radio_Astronomy}
Kraus, J. 1986, Radio Astronomy, 2nd edn. (Powell, Ohio USA: Cygnus-Quasar
  Books)

\bibitem[{{Larson} {et~al.}(2010){Larson}, {Dunkley}, {Hinshaw}, {Komatsu},
  {Nolta}, {Bennett}, {Gold}, {Halpern}, {Hill}, {Jarosik}, {Kogut}, {Limon},
  {Meyer}, {Odegard}, {Page}, {Smith}, {Spergel}, {Tucker}, {Weiland},
  {Wollack}, \& {Wright}}]{WMAP7:powerspectra}
{Larson}, D., {Dunkley}, J., {Hinshaw}, G., {et~al.} 2010

\bibitem[{Leitch {et~al.}(2005)Leitch, Kovac, Halverson, Carlstrom, Pryke, \&
  Smith}]{DASI:3yr}
Leitch, E.~M., Kovac, J.~M., Halverson, N.~W., {et~al.} 2005, \apj, 624, 10

\bibitem[{Leitch {et~al.}(2002)Leitch, Kovac, Pryke, Reddall, Sandberg,
  Dragovan, Carlstrom, Halverson, \& Holzapfel}]{DASI:instrument}
Leitch, E.~M., Kovac, J.~M., Pryke, C., {et~al.} 2002, Nature, 420, 763

\bibitem[{{Lewis}(2003)}]{EB_harmonic}
{Lewis}, A. 2003, \prd, 68, 083509

\bibitem[{{Lewis} {et~al.}(2002){Lewis}, {Challinor}, \&
  {Turok}}]{EB_incomplete_sky}
{Lewis}, A., {Challinor}, A., \& {Turok}, N. 2002, \prd, 65, 023505

\bibitem[{Liddle \& Lyth(2000)}]{Inflation}
Liddle, A.~R. \& Lyth, D.~H. 2000, Cosmological Inflation and Large-Scale
  Structure, 1st edn. (Cambridge University Press)

\bibitem[{Mukhanov(2005)}]{Foundations_Cosmology}
Mukhanov, V. 2005, Physical Foundations of Cosmology, 1st edn. (Cambridge
  University Press)

\bibitem[{{Ng} \& {Liu}(1999)}]{Ng_correlation_pol}
{Ng}, K. \& {Liu}, G. 1999, International Journal of Modern Physics D, 8, 61

\bibitem[{Nolta \& et~al.(2008)}]{WMAP5:powerspectra}
Nolta, M.~R. \& et~al. 2008, submitted to {\apj}, arXiv:0803.0593

\bibitem[{Pryke \& et~al.(2009)}]{QUaD2}
Pryke, C. \& et~al. 2009, \apj, 692, 1247

\bibitem[{Pryke {et~al.}(2002{\natexlab{a}})Pryke, Halverson, Kovac, Davidson,
  LaRoque, Schartman, Yamasaki, Carlstrom, Holzapfel, Dragovan, Cartwright,
  Mason, Padin, Pearson, Shepherd, \& Readhead}]{DASI:I}
Pryke, C., Halverson, N.~W., Kovac, J.~M., {et~al.} 2002{\natexlab{a}}, \apj,
  568, 28

\bibitem[{Pryke {et~al.}(2002{\natexlab{b}})Pryke, Halverson, Leitch, Kovac,
  Carlstrom, Holzapfel, \& Dragovan}]{DASI:III}
Pryke, C., Halverson, N.~W., Leitch, E.~M., {et~al.} 2002{\natexlab{b}}, \apj,
  568, 46

\bibitem[{Rohlfs \& Wilson(2003)}]{Tools_Radio_Astronomy}
Rohlfs, K. \& Wilson, T.~L. 2003, Tools of Radio Astronomy, 4th edn. (New York,
  NY USA: Springer-Verlag)

\bibitem[{{Seljak} \& {Hirata}(2004)}]{B_lensing}
{Seljak}, U. \& {Hirata}, C.~M. 2004, \prd, 69, 043005

\bibitem[{{Smith}(2006)}]{Smith:pseudo_EB}
{Smith}, K.~M. 2006, \prd, 74, 083002

\bibitem[{{The Planck Collaboration}(2006)}]{Planck_bluebook}
{The Planck Collaboration}. 2006, ArXiv Astrophysics e-prints

\bibitem[{{Tucci} {et~al.}(2005){Tucci}, {Mart{\'{\i}}nez-Gonz{\'a}lez},
  {Vielva}, \& {Delabrouille}}]{B_mode_limit_foreground}
{Tucci}, M., {Mart{\'{\i}}nez-Gonz{\'a}lez}, E., {Vielva}, P., \&
  {Delabrouille}, J. 2005, \mnras, 360, 935

\bibitem[{{Wandelt} {et~al.}(2001){Wandelt}, {Hivon}, \&
  {G{\'o}rski}}]{pseudo_Cl}
{Wandelt}, B.~D., {Hivon}, E., \& {G{\'o}rski}, K.~M. 2001, \prd, 64, 083003

\bibitem[{Zaldarriaga(1998)}]{Zaldarriaga:Polarization_Exp}
Zaldarriaga, M. 1998, \apj, 503, 1

\bibitem[{Zaldarriaga \& Seljak(1997)}]{Seljak-Zaldarriaga:Polarization}
Zaldarriaga, M. \& Seljak, U. 1997, \prd, 55, 1830

\end{thebibliography}

\end{document}